\documentclass[twocolumn,letter]{jpsj2} %% two-column layout
\def\runauthor{Mamoru {\sc Yogi}}
\title{Novel Phase Transition Near the Quantum Critical Point in the Filled-Skutterudite Compound CeOs$_{4}$Sb$_{12}$: an Sb-NQR Study}

\author{
Mamoru \textsc{Yogi}$^{1}$\thanks{E-mail: yogi@nmr.mp.es.osaka-u.ac.jp}\thanks{Present address: Faculty of Science, University of the Ryukyus, Okinawa 903-0213},
Hisashi \textsc{Kotegawa}$^{1}$\thanks{Present address: Department of Physics, Faculty of Science, Okayama University, Okayama 700-8530},
Guo-qing \textsc{Zheng}$^{1 \ddagger}$,
Yoshio \textsc{Kitaoka}$^{1}$,
Shuji \textsc{Ohsaki}$^{2}$,
Hitoshi \textsc{Sugawara}$^{2}$\thanks{Present address: Department of Physics, Faculty of Integrated Arts and Sciences, Tokushima University, Tokushima 770},
and Hideyuki \textsc{Sato}$^{2}$
}

\inst{
$^{1}$Department of Materials Engineering Science, Graduate School of Engineering Science, Osaka University,  Osaka 560-8531, Japan\\
$^{2}$Graduate School of Science, Tokyo Metropolitan University, Minami-Ohsawa 1-1, Hachioji, Tokyo 192-0397, Japan
}

\abst{
We report on a novel phase transition at $T$ = 0.9 K in the Ce-based filled-skutterudite compound CeOs$_{4}$Sb$_{12}$ via measurements of the nuclear-spin lattice relaxation rate $1/T_{1}$ and nuclear quadrupole resonance (NQR) spectrum of Sb nuclei.
The temperature ($T$) dependence of $1/T_1$ behaves as if approaching closely an antiferromagnetic (AFM) quantum critical point (QCP), following the relation $1/T_{1}\propto T/(T-T_{\rm N})^{1/2}$ with $T_{\rm N} = 0.06$ K in the range of $T=1.3-25$ K.
The onset of either the spin-density-wave (SDW) or charge-density-wave (CDW) order at $T_{\rm 0} = 0.9$ K, that is, of the first order, is evidenced by a broadening of the NQR spectrum and a marked reduction in $1/T_1$ just below $T_{\rm 0}$. The $f$-electron-derived correlated band realized in CeOs$_{4}$Sb$_{12}$ is demonstrated to give rise to the novel phase transition on the verge of AFM QCP.
}

\kword{filled skutterudite, Kondo semiconductor, quantum critical point, SDW/CDW, NQR, CeOs$_{4}$Sb$_{12}$}

\begin{document}
\maketitle
\newpage

Filled-skutterudite compounds ReT$_4$Pn$_{12}$ (Re = rare earth; T = Fe, Ru or Os; Pn = P, As or Sb) have attracted much attention because of their rich physical phenomena that are not yet fully understood.
For instance, PrFe$_4$P$_{12}$ shows a heavy fermion (HF)-like behavior with a large mass of $m^*\sim 70~m_e$  under a magnetic field ($H$) \cite{Sugawara} and undergoes an anomalous transition at a temperature $T = 6.4$ K at $H = 0$, indicative of a quadrupolar ordering \cite{Matsuda,Aoki,Nakanishi}.
Note that the large $C/T$ value for this compound is due not only to low-energy degrees of freedom such as either magnetic or quadrupolar fluctuations, but also to the Schottky anomaly originating from some low-lying CEF splitting.
PrOs$_4$Sb$_{12}$ is the first Pr-based HF superconductor to reveal the large jump in the specific heat at $T_c=1.85$ K, the slope of the upper critical field near $T_c$, and the electronic specific heat coefficient $\gamma\sim 350-500$ mJ/K$^2$mol in the normal state \cite{Bauer_PrOsSb}.
The CEF-energy scheme of PrOs$_{4}$Sb$_{12}$ has recently been determined to be the singlet $\Gamma_{1}$ ground state accompanying the very low lying excited state $\Gamma_{4}^{(2)}$ \cite{Kohgi, Maple}, and it was suggested that the HF-like behavior exhibited by PrOs$_4$Sb$_{12}$ may be relevant to a very close energy level between $\Gamma_{1}$ and $\Gamma_{4}$.
Thus, in the Pr-based compound, the 4$f^2$-derived CEF effect plays an important role in their rich physical phenomena.

In contrast, most Ce-based filled-skutterudite compounds show semiconducting behavior, on the basis of which they are called {\it hybridization-gap semiconductors}.
A series of CeT$_{4}$P$_{12}$ (T = Fe, Ru or Os) compounds have a hybridization gap of 400$\sim$1500 K, and as the lattice constant increases, the energy gap decreases \cite{Meisner, Shirotani}.
CeT$_{4}$Sb$_{12}$ compounds show semimetallic behavior. CeOs$_{4}$Sb$_{12}$ is, on the one hand, suggested to exhibit Kondo insulating behavior with a large specific heat coefficient, $\gamma \sim 92$ mJ/K$^2$mol, and a very small gap of about $\Delta/k_{\rm B}\sim 10$ K at the Fermi level\cite{Bauer}.
Filled-skutterudite compounds form a unique cubic structure (space group: $Im$\={3}, $T_{h}^{5}$ No.204).
The rare-earth ion in this structure is surrounded by twelve Sb atoms that are strongly hybridized with nearly localized $4f$ electrons.
Band calculations on CeOs$_{4}$Sb$_{12}$ revealed a unique band structure where no band gap is formed at the Fermi level \cite{Harima}.
Electrical-resistivity measurement under pressure suggests that a hopping conductivity mechanism is realized at low temperatures \cite{Hedo}.
Furthermore, CeOs$_{4}$Sb$_{12}$ was reported to exhibit an anomaly at approximately $T = 1$ K.
The entropy release below this temperature is not sufficiently large to be considered to arise from some intrinsic phase transition, but is indicative of some impurity effect \cite{Bauer}.
On the other hand, measurements of specific heat, magnetoresistance and the Hall effect under $H$ revealed that a phase transition takes place \cite{Namiki,Sugawara_CeOsSb}.

Here, we report on a novel phase transition in CeOs$_{4}$Sb$_{12}$ that emerges at $T$ = 0.9 K on the verge of an antiferromagnetic (AFM) quantum critical point (QCP) via measurements of the nuclear-spin lattice relaxation rate $1/T_{1}$ and nuclear quadrupole resonance (NQR) spectrum of Sb nuclei.

Single crystals of CeOs$_{4}$Sb$_{12}$ were grown by the Sb-flux method \cite{Namiki}.
The Sb-NQR measurement on the single crystal, which was crushed into powder, was performed using the conventional spin-echo method at $H$ = 0 and in the range of $T = 0.2-300$ K using a $^3$He-$^{4}$He-dilution refrigerator.

\begin{figure}[tb]
  \begin{center}
    \includegraphics[keepaspectratio=true,height=45mm]{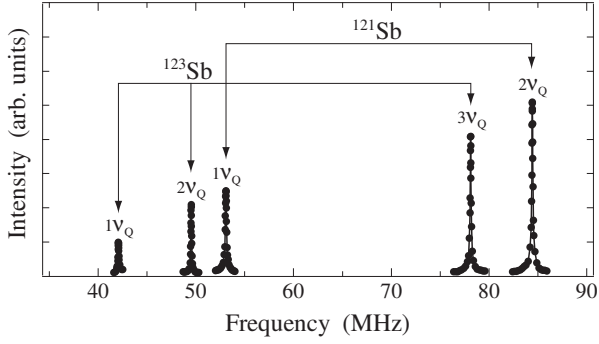}
  \end{center}
  \caption{$^{121}$Sb- and $^{123}$Sb-NQR spectra for CeOs$_{4}$Sb$_{12}$ at $T$ = 4.2 K.}
\end{figure}

For CeOs$_{4}$Sb$_{12}$, five NQR transitions at 4.2 K are shown in Fig.1.
These five NQR transitions come from the two Sb isotopes $^{121}$Sb and $^{123}$Sb with natural abundances 57.3\% and 42.7\% and nuclear spins $I$ = 5/2 and 7/2, respectively, and with a nuclear quadrupole moment ratio of $^{123}Q/^{121}Q \sim$ 1.361.
Thus, two and three NQR transitions are observed for $^{121}$Sb and $^{123}$Sb, respectively \cite{NMR_Handbook}.
The NQR Hamiltonian is described as
\begin{equation}
{\cal H}_{Q}=\frac{h\nu_{Q}}{6}[3I_{z}^{2}-I(I+1)+\frac{\eta}{2}(I_{+}^{2}+I_{-}^{2})],
\end{equation}
where $I$ is the nuclear spin, $\nu_{Q} \equiv 3e^{2}qQ/2I(2I-1)h$ is the nuclear quadrupole frequency, and $\eta$ is the asymmetry parameter defined as $\eta=(V_{xx}-V_{yy})/V_{zz}$.
Here, $V_{xx}$, $V_{yy}$, and $V_{zz}$ are the components of the electric field gradient (EFG) tensor.
The values of respective $^{121}$Sb and $^{123}$Sb NQR frequencies, $^{121}\nu_{Q}\sim$ 43.861 MHz and $^{123}\nu_{Q}\sim$ 26.630 MHz, are estimated together with $\eta\sim 0.463$.
These two NQR frequencies are due to the difference in nuclear quadrupole moment $Q$, i.e., $^{123}Q/^{121}Q \sim$ 1.361. 
CeOs$_{4}$Sb$_{12}$ and PrOs$_{4}$Sb$_{12}$ have the same crystal structure.
It is then expected that the values of $\nu_{Q}$ and $\eta$ are almost equivalent to those in PrOs$_{4}$Sb$_{12}$ reported previously \cite{Kotegawa}.
The full width at half maximum (FWHM) $\Delta f$ in the NQR spectrum is as small as $\sim$ 65 kHz at 4.2 K, indicating the high quality of the sample.
\begin{figure}[tb]
  \begin{center}
    \includegraphics[keepaspectratio=true,height=75mm]{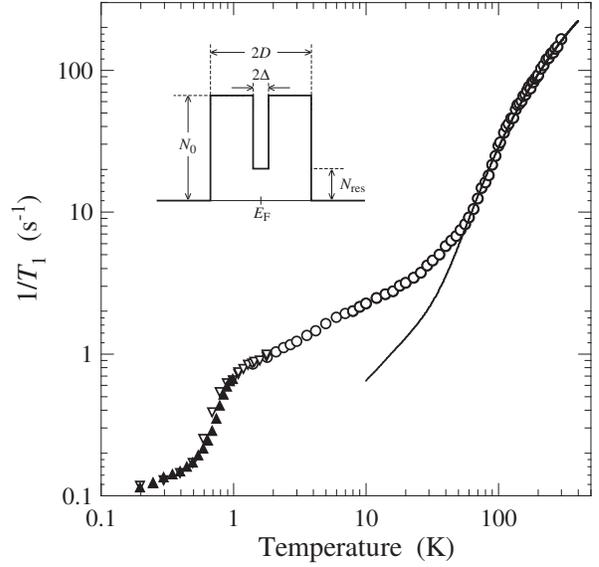}
  \end{center}
  \caption{The temperature ($T$) dependence of $1/T_1$ for $^{123}$Sb-$2\nu_Q$ transition at $H$ = 0. The open (solid) triangles show $1/T_{1}$ in the $T$ decrease (increase) process. $1/T_{1}$ does not show the $T$ hysteresis above 1.4 K (open circles). The solid line is the $T$ dependence of $1/T_{1}$ calculated by assuming a rectangular shape of density of states (DOS) with a bandwidth 2$D = 3000$ K, a gap size of 2$\Delta = 320$ K and an amount of residual DOS $N_{\rm 0}/N_{\rm res} = 0.3$ (see text) as shown in the inset.}
\end{figure}

$T_{1}$ is measured for both the $^{123}$Sb-$2\nu_Q$ ($\pm 3/2\leftrightarrow \pm 5/2$) and $^{121}$Sb-$2\nu_Q$ transitions above 1.4 K.
For NQR-$T_1$, note that either a magnetic or a quadrupole relaxation process or both are possible as its origin.
In case of the magnetic process, $1/T_{1}$ should be proportional to the square of the nuclear gyromagnetic ratio ($\gamma_{\rm N}^{2}$) that is scaled to the nuclear magnetic moment.
This is the present case because the values of $(1/T_{1})/{\gamma_{\rm N}}^{2}$ are equivalent for $^{121}$Sb and $^{123}$Sb, where $^{121}\gamma_{\rm N}= 10.2549$ and $^{123}\gamma_{\rm N}=5.5530$ MHz/T, respectively \cite{NMR_Handbook}.
$T_1$ was uniquely determined by a theoretical curve for the recovery of nuclear magnetization $m(t)$ in which the asymmetry parameters were incorporated \cite{Chepin}.
In the case of $I=7/2$ and $\eta = 0.46$, the recovery curve is given by
\begin{eqnarray}
&&m_{2\nu_Q}(t)=\frac{M(\infty )-M(t)}{M(\infty )}\nonumber \\
&&= 0.077\exp\left(\frac{-3t}{T_{1}} \right)+0.0165\exp\left(\frac{-8.562t}{T_{1}} \right)\nonumber \\
&&+0.9065\exp \left(\frac{-17.207t}{T_{1}} \right).
\end{eqnarray}
The $T$ dependence of $1/T_{1}$ for $^{123}$Sb-2$\nu_{Q}$ is indicated in Fig. 2.
The $1/T_{1}$ in the range of $T = 200-300$ K decreases  exponentially, and then $1/T_{1}$ is well fitted by an activation type of formula $1/T_{1}\sim \exp(-\Delta/k_{\rm B}T)$ with a gap value of $\Delta/k_{\rm B}\sim 330$ K.
The gap value coincides with the energy separation $\Delta/k_{B}\sim 327$ K between the CEF ground state $\Gamma_{7}$ and the excited state $\Gamma_{8}$.
Here, $\Delta/k_{B}\sim 327$ K is cited from a measurement of magnetic susceptibility \cite{Bauer}.
Such $1/T_{1}$ is, however, observed in the related compound CeRu$_{4}$Sb$_{12}$, where the magnetic susceptibility measurement does not reveal a clear CEF effect \cite{Yogi_CeRuSb,Bauer_CeRuSb}.
Also, recent ultrasound measurement has not detected a clear CEF effect on CeOs$_{4}$Sb$_{12}$ \cite{Nakanishi2}.
Therefore, this exponential decrease in $1/T_1$ may not be related to the CEF effect.
In heavy-fermion systems, $1/T_{1}$ is dominated by $f$-electron-derived spin fluctuations through the hybridization between $f$ and conduction electrons.

In general, $1/T_{1}$ is given by \cite{Moriya1}
\begin{equation}
\frac{1}{T_{1}} = \frac{2\gamma_{\rm n}^{2}k_{\rm B}T}{(\gamma_{\rm e}\hbar)^2}\sum_{\boldsymbol{q}}A_{\boldsymbol{q}}A_{-\boldsymbol{q}}\frac{\chi_{\perp}^{''}(\boldsymbol{q}, \omega_{0})}{\omega_{0}},
\end{equation}
where $A_{q}$ is the $q$-dependent hyperfine coupling constant, and $\chi_{\perp}^{''}$ is the perpendicular component of the imaginary part of dynamical susceptibility.
Provided that the $q$ dependence of spin fluctuations is not so large, eq. (3) can be rewritten as 
\begin{equation}
\frac{1}{T_{1}}\propto T\int N_{eff}^{2}(E)\left\{-\frac{\partial f(E)}{\partial E}\right\}dE,  
\end{equation}
where $f(E)$ is the Fermi distribution function and $N_{eff}(E)$ is the renormalized quasiparticle density of state (DOS) at the Fermi level.
For "Kondo" insulators reported to date, $1/T_{1}$ is described by a pseudogap model induced by the hybridization between $f$ and conduction electrons.
In CeRhSb and CeNiSb, which form orthorhombic structures, $1/T_{1}$ is explained by the V-shaped gap model \cite{Ce111_1, Ce111_2}.
In Ce$_{3}$Bi$_{4}$Pt$_{3}$, which forms a cubic structure, $1/T_{1}$ that decreases exponentially is explained by assuming an isotropic gap model with a rectangular DOS \cite{CeBiPt}.
The difference in the gap structure for the "Kondo" insulators may be related to their crystal structure.
In this context, for CeOs$_{4}$Sb$_{12}$, which forms a cubic structure, the isotropic gap model with a rectangular DOS is expected to be applied for interpreting $1/T_1$. 
The solid line in Fig. 2 shows  $1/T_{1}$ calculated by the model DOS illustrated in the inset. Here, the fitting parameters are $2D = 3000$ K and $2\Delta = 320$ K with a finite residual DOS $N_{\rm 0}/N_{\rm res} = 0.3$ at the Fermi level. This model explains the $1/T_{1}$ above $T\sim$ 90 K.
\begin{figure}[tb]
  \begin{center}
    \includegraphics[keepaspectratio=true,height=75mm]{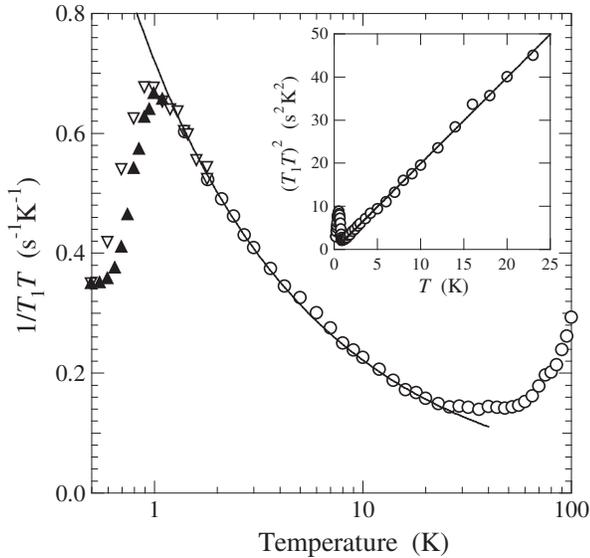}
  \end{center}
  \caption{The $T$ dependence of $1/T_{1}T$ where open circles correspond to the data of $1/T_{1}$ (open circles) in Fig. 2. The data in the range of $T = 1.3-20$ K are fitted by the relation $1/T_{1}\propto T/(T-T_{\rm N})^{1/2}$ with $T_{\rm N}= 0.06$ K as shown by the solid curve. Note that this relation is consistent with the SCR theory for three-dimensional itinerant weakly antiferromagnetic metals \cite{Moriya2,Ishigaki,Nakamura}  The inset shows a $^{123}(T_{1}T)^2$ vs $T$ plot where the solid line corresponds to the relation $(T_1T)^2 \propto (T-T_{\rm N})$, ensuring the theoretical prediction in terms of the SCR theory mentioned above.}
\end{figure}
If the residual DOS at the Fermi level existing inside the pseudogap is responsible for the relaxation process at a low $T$, a behavior in which $T_{1}T$ = constant would be expected at temperatures lower than $\Delta/k_B$.
Actually, although a behavior in which $T_{1}T$ = constant is observed, such behavior is only valid in the range $T = 25-50$ K.
Unexpectedly, $1/T_{1}T$ increases upon cooling below 25 K as shown in Fig. 3.
This enhancement in $1/T_{1}T$ has not been observed for other "Kondo" insulators.
It is remarkable that $1/T_{1}$ behaves as $1/T_{1} \propto \sqrt{T}$ below $T = 25$ K.
The self-consistent renormalization (SCR) theory for spin fluctuations in itinerant AFM metals has been applied to the case of HF systems near an AFM QCP \cite{Moriya2,Ishigaki,Nakamura}.
In this case, it was predicted that $1/T_{1}$ obeys the relation
\begin{equation}
\frac{1}{T_{1}} \propto T\sqrt{\chi_Q(T)} = \frac{T}{\sqrt{T-T_{\rm N}}}
\end{equation}
because of the Currie-Weiss law of staggered susceptibility $\chi_Q(T)\propto 1/(T-T_{\rm N})$ with an AFM ordering temperature $T_{\rm N}$.
Therefore, note that $1/T_{1}\propto \sqrt{T}$ is valid at $T_{\rm N}\sim 0$ in the vicinity of an AFM QCP.
In order to further inspect this unique $T$ dependence,  $(T_{1}T)^{2}$ is plotted as a function of $T$ in the inset of Fig. 3.
From this plot, it is ensured that the relation $(T_{1}T)^{2}\propto (T-T_{\rm N})$ is valid with $T_{\rm N} = 0.06$ K in the range $T = 1.3- 25$ K.
It is interesting that CeOs$_4$Sb$_{12}$ is near the AFM QCP and an estimation of "$T_{\rm N} = 0.06$ K" in $\chi_Q(T)$ is indicative of a possible onset of some AFM order at low temperatures.
\begin{figure}[tb]
  \begin{center}
    \includegraphics[keepaspectratio=true,height=55mm]{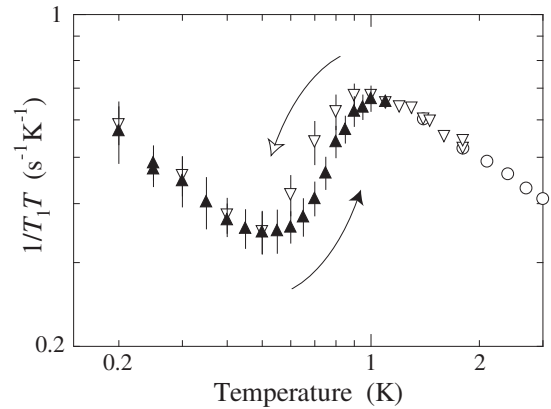}
  \end{center}
  \caption{The $T$ dependence of $1/T_{1}T$ at low temperatures. Open and closed triangles show the $T$ dependence of $1/T_{1}T$ upon cooling and heating, respectively.}
\end{figure}
As a matter of fact, suggesting the appearance of some anomaly, the recovery curve below 1.4 K is not described by eq. (1).
Therefore, a short component in $T_1$ is plotted below 1.4 K in Fig. 4.
With further decreasing $T$, a marked decrease in $1/T_1$ is observed below $T = 0.9$ K.
Figure 4 shows the detailed $T$ dependence of $1/T_{1}T$ around $T = 0.9$ K.
Unexpectedly, this $T_1$ result reveals the onset of some phase transition at $T$=0.9 K, which is higher than "$T_{\rm N} = 0.06$ K" estimated from $\chi_Q(T)$.
The sudden decrease in $1/T_{1}T$ is due to the opening of a gap at the Fermi surface, caused by the onset of phase transition.
The data are consistent with such an activation type of behavior as $1/T_{1}\propto$ exp$(-\Delta/k_{\rm B}T)$ with a gap size of $\Delta/k_{\rm B}\sim 1.83$ K.
It should be noted that $1/T_{1}T$ undergoes clear hysteresis upon cooling and heating.
The hysteresis in $1/T_{1}T$ suggests a first-order transition emerging at $T = 0.9$ K, which is also corroborated by the measurement of the $T$ dependence of the NQR spectral shape as shown later.
\begin{figure}[tb]
  \begin{center}
    \includegraphics[keepaspectratio=true,height=55mm]{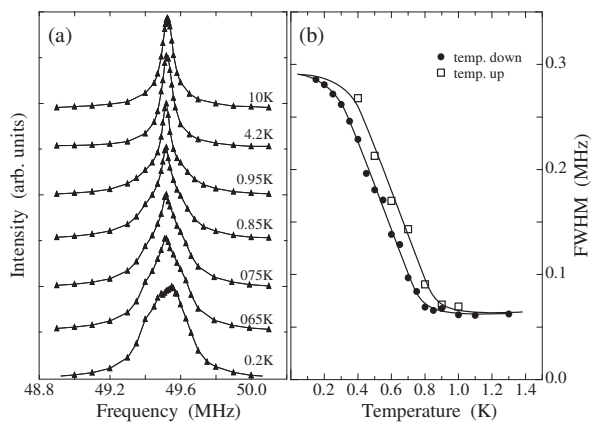}
  \end{center}
  \caption{(a) The $T$ dependence of the NQR spectrum of $^{123}$Sb-$2\nu_Q$ transitions. (b) The $T$ dependence of FWHM upon cooling (closed circles) and heating (open squares).}
\end{figure}

Figure 5(a) indicates the $T$ dependence of the NQR spectrum at the $^{123}$Sb-2$\nu_Q$ transition below 4.2 K. Above 0.9 K, it exhibits a Lorentzian shape with a small value of $\Delta f = 65$ kHz.
With decreasing $T$ below 0.9 K, it starts to broaden as clearly seen in the figure, and at the same time, $1/T_{1}$ is markedly reduced.
Note that the tail in the NQR spectrum is significantly increased and makes the NQR spectral shape deviate from its Lorentzian form.
This result may be ascribed to either the appearance of an internal field that makes its magnitude distributed at the Sb site or a distribution of $\nu_{\rm Q}$.
Then, such a characteristic spectral shape below $T = 0.9$ K points to either a wide distribution of the internal field at Sb if a transition were a spin density wave (SDW), or a distribution in the electric field gradient associated with a charge density wave (CDW), both presumably triggered by the nesting effect of the Fermi surface.
Furthermore, the $T$ dependence of FWHM in the NQR spectrum shows hysteresis upon cooling and heating as shown in Fig. 5(b).
This result also suggests that the first-order transition begins to take place below $T_{\rm 0} = 0.9$ K, consistent with the hysteresis observed in the $T$ dependence in $1/T_1$.
The present results on CeOs$_4$Sb$_{12}$ have revealed that the phase transition near the AFM QCP is of the first order.
This finding deserves a theoretical study on the physics behind the AFM QCP in $f$-electron-derived correlated bands realized in HF systems in general.

In summary, we have shown from the measurements of $1/T_{1}$ and the NQR spectrum of Sb nuclei that CeOs$_{4}$Sb$_{12}$ is a unique Kondo semiconductor.
CeOs$_{4}$Sb$_{12}$ has a hybridization gap similar to other Kondo semiconductors, however, there exists a large residual DOS within the gap.
The enhancement of spin fluctuations below 20 K indicates that CeOs$_{4}$Sb$_{12}$ is closely located to the AFM QCP.
Even in such a case, the remarkable finding is that either a SDW or CDW order, which is of the first order, takes place at $T = 0.9$ K, possibly triggered by the nesting effect of the Fermi surface.
The present result, we believe, deserves a theoretical study on the physics behind the AFM QCP realized in HF systems in general.

{\it note added in proof}: 
After the submission of this paper, we became aware of an important theoretical work by T. Ohashi, A. Koga, S. Suga, and N. Kawakami (Phys. Rev. B70 (2004) 245104). This work has suggested that if some state were induced inside the Kondo gap, the antiferromagnetic correlation might be developed, leading to an SDW instability at low temperatures as observed in the present study. 

\section*{Acknowledgment}
We wish to thank H. Harima and K. Ishida for their helpful and valuable comments.
This work was partially supported by a Grant-in-Aid for Creative Scientific Research (15GS0213), MEXT, The 21st Century COE Program supported by the Japan Society for the Promotion of Science, and a Grant-in-Aid for Scientific Research in the Priority Area "Skutterudite" (No. 15072206), MEXT.


\begin{thebibliography}{99} %% The number "99" means that this list has more than nine items.
\bibitem{Sugawara} H. Sugawara, T. D. Matsuda, K. Abe, Y. Aoki, H. Sato, S. Nojiri, Y. Inada, R. Settai and Y.~\=Onuki: Phys. Rev. B {\textbf 66} (2002) 134411.
\bibitem{Matsuda} T. D. Matsuda, H. Okada, H. Sugawara, Y. Aoki, H. Sato, A. V. Andreev, Y. Shiokawa, V. Sechovsky, T. Honma, E. Yamamoto and Y. Onuki: Physica B {\textbf 281-282} (2000) 220.
\bibitem{Aoki} Y. Aoki, T. Namiki, T. D. Matsuda, K. Abe, H. Sugawara and H. Sato: Phys. Rev. B {\textbf 65} (2002) 064446.
\bibitem{Nakanishi} Y. Nakanishi, T. Simizu, M. Yoshizawa, T. D. Matsuda, H. Sugawara and H. Sato: Phys. Rev. B {\textbf 63} (2002) 184429.
\bibitem{Bauer_PrOsSb} E. D. Bauer, N. A. Frederick, P.-C. Ho, V. S. Zapf and M. B. Maple: Phys. Rev. B {\textbf 65} (2002) 100506(R).
\bibitem{Kohgi} M. Kohgi, K. Iwasa, M. Nakajima, N. Metoki, S. Araki, N. Bernhoeft, J.-M. Mignot, A. Gukasov, H. Sato, Y. Aoki and H. Sugawara: J.\ Phys.\ Soc.\ Jpn.\ {\textbf 72} (2003) 1002.
\bibitem{Maple} E. A. Goremychkin, R. Osborn, E. D. Bauer, M. B. Maple, N. A. Frederick, W. M. Yuhasz, F. M. Woodward and J. W. Lynn: Phys. Rev. Lett. {\textbf 93} (2004) 157003.
\bibitem{Meisner} G. P. Meisner, M. S. Torikachvili, K. N. Yang, M. B. Maple and R. P. Guertin: J. Appl. Phys. {\textbf 57} (1985) 3073.
\bibitem{Shirotani} I. Shirotani, T. Uchiumi, C.Sekine, M. Hori, S. Kimura and N. Hamaya: J. Solid State Chem. {\textbf 142} (1999) 146.
\bibitem{Bauer} E. D. Bauer, A. \'Slebarski, E. J. Freeman, C. Sirvent and M. B. Maple:  J. Phys.: Condens. Matter {\textbf 13} (2001) 4495.
\bibitem{Harima} H. Harima and K. Takegahara: J. Phys.: Condens. Matter {\textbf 15} (2003) S2081-S2086.
\bibitem{Hedo} M. Hedo, Y. Uwatoko, H. Sugawara and H. Sato: Physica B {\textbf 329-333} (2003) 456.
\bibitem{Namiki} T. Namiki, Y. Aoki, H. Sugawara and H. Sato: Acta Phys. Pol. B, {\textbf 34} (2003) 1161.
\bibitem{Sugawara_CeOsSb} H. Sugawara, S. Osaki, M. Kobayashi, T. Namiki, S.R. Saha, Y. Aoki and H. Sato: Phys. Rev. B {\textbf 71} (2005) 125127.
\bibitem{NMR_Handbook} CRC Handbook of Chemistry and Physics, 82nd Ed. (CRC Press, Boca Raton, FL, 2001).
\bibitem{Kotegawa} H. Kotegawa, M. Yogi, Y. Imamura, Y. Kawasaki, G.-q. Zheng, Y. Kitaoka, S. Ohsaki, H. Sugawara, Y. Aoki and H. Sato: Phys. Rev. Lett. {\textbf 90} (2003) 027001.
\bibitem{Chepin} J. Chepin and J. H. Ross, Jr.: J. Phys.: Condens. Matter {\textbf 3} (1991) 8103.
\bibitem{Yogi_CeRuSb} M. Yogi {\it et al.}: unpublished.
\bibitem{Bauer_CeRuSb} E. D. Bauer, A. \'Slebarski, R. P. Dickey, E. J. Freeman, C. Sirvent, V. S. Zapf, N. R. Dilley and M. B. Maple: J. Phys.: Condens. Matter {\textbf 13} (2001) 5183.
\bibitem{Nakanishi2} Y. Nakanishi, M. Oikawa, T. Kumagai, M. Yoshizawa, T. Namiki, H. Sugawara and H. Sato: to be published in Physica B.
\bibitem{Moriya1} T.~Moriya: J.\ Phys.\ Soc.\ Jpn. {\textbf 18} (1963) 516.
\bibitem{Ce111_1} K. Nakamura, Y. Kitaoka, K. Asayama, T. Takabatake, H. Tanaka and H. Fujii: J.\ Phys.\ Soc.\ Jpn. {\textbf 63} (1994) 433.
\bibitem{Ce111_2} K. Nakamura, Y. Kitaoka, K. Asayama, T. Takabatake, G. Nakamoto, H. Tanaka and H. Fujii: Phys. Rev. B {\textbf 53} (1996) 6385.
\bibitem{CeBiPt} A. P. Reyes, R. H. Heffner, P. C. Canfield, J. D. Thompson and Z. Fisk: Phys. Rev. B {\textbf 49} (1994) 16321.
\bibitem{Moriya2} T.~Moriya and T.~Takimoto: J.\ Phys.\ Soc.\ Jpn. {\textbf 64} (1995) 960.
\bibitem{Ishigaki} A.~Ishigaki and T.~Moriya: J.\ Phys.\ Soc.\ Jpn.\ {\textbf 65} (1996) 3402.
\bibitem{Nakamura} S.~Nakamura, T.~Moriya and K.~Ueda: J.\ Phys.\ Soc.\ Jpn. {\textbf 65} (1996) 4026.
\end{thebibliography}
\end{document}